\documentclass[
aps,     
prl,                    
showpacs,               
superscriptaddress,     
nofootinbib,            
twocolumn,              %
floatfix,
nobalancelastpage
]{revtex4}
\usepackage{graphicx}
\usepackage{epstopdf}
\usepackage{natbib}
\begin{document}


\title{\large
Stringent Constraint on Galactic Positron Production
\normalsize}

\author{John F. Beacom}
\affiliation{Department of Physics, The Ohio State University, Columbus, Ohio
43210, USA}
\affiliation{Department of Astronomy, The Ohio State University, Columbus, Ohio
43210, USA}
\author{Hasan Y{\"u}ksel}
\affiliation{Department of Physics, The Ohio State University, Columbus, Ohio
43210, USA}

\date{15 December 2005; minor revisions, 24 July 2006}

\begin{abstract}
The intense 0.511 MeV gamma-ray line emission from the Galactic Center observed by INTEGRAL requires a large annihilation rate of nonrelativistic positrons.  If these positrons are injected at even mildly relativistic energies, higher-energy gamma rays will also be produced.   We calculate the gamma-ray spectrum due to inflight annihilation and compare to the observed diffuse Galactic gamma-ray data.  Even in a simplified but conservative treatment, we find that the positron injection energies must be $\lesssim 3$ MeV, which strongly constrains models for Galactic  positron production.
\end{abstract}



\maketitle


The central region of the Milky Way Galaxy (hereafter the GC) is illuminated by the annihilation of $\sim 10^{50}$ positrons per year, producing the 0.511 MeV gamma-ray line flux of $\left(1.07 \pm 0.03 \right) \times 10^{-3}$ photons cm$^{-2}$ s$^{-1}$ that is robustly detected by the INTEGRAL satellite \cite{INTEGRALearly, Churazov04, Knodlseder05, Jean05, Weidenspointner06, Guessoum05, Strong05}.  The angular distribution of this emission is aligned with the GC and is consistent with a 2-dimensional Gaussian of $\simeq 8^\circ$ FWHM.  There is only weak evidence for a disk component, in stark contrast to gamma-ray maps tracing nucleosynthesis (e.g., the 1.809 MeV line from decaying $^{26}$Al) or cosmic ray processes (e.g., the 1--30 MeV continuum), which reveal a bright Galactic disk with several hot regions, and a much less prominent central region~\cite{Knodlseder99,Strong98}.  Both the flux and angular distribution of the 0.511 MeV radiation are difficult to explain~\cite{INTEGRALearly, Churazov04, Knodlseder05, Jean05, Weidenspointner06, Guessoum05, Strong05, Models-Astro, Models-Exotic}.

Central to resolving the origin of the positrons is the question of their injection energies, which range up to 100 MeV or even higher in recent astrophysical~\cite{Models-Astro} and exotic (requiring new particle physics)~\cite{Models-Exotic} models.  However, after energy loss (and diffusion and delay), the positron annihilations produce only gamma rays {\it at or below} 0.511 MeV, concealing their true injection energies (and production sites).  To circumvent this, we must consider the relatively rare gamma-ray emission {\it above} 0.511 MeV produced by relativistic positrons.  The Internal Bremsstrahlung (IB) radiation associated with positron production (a Q.E.D.~radiative correction)  would conflict with COMPTEL and EGRET diffuse gamma-ray observations unless the injection energy is $\lesssim 20$ MeV~\cite{Beacom04}.

Here we consider the gamma rays produced by the Inflight Annihilation (IA) of energetic positrons with electrons in the interstellar medium; as noted long ago by Heitler, up to $\sim 20\%$ of relativistic positrons annihilate in flight while undergoing ionization energy loss in matter~\cite{Heitler}.  The IA radiation as a probe of astrophysical positrons has also been used in Refs.~\cite{Stecker, Furlanetto, Murphy, Svensson, Aharonian, Moskalenko}, with Ref.~\cite{Aharonian} constraining the Galactic positrons using earlier data.   The new, high-quality data from INTEGRAL allow us to model-independently study the Galactic positrons, normalizing the intensity and angular distribution of the IA radiation to the 0.511 MeV line emission, and arriving at a very stringent limit.

\begin{figure}
\includegraphics[width=3.25in,clip=true]{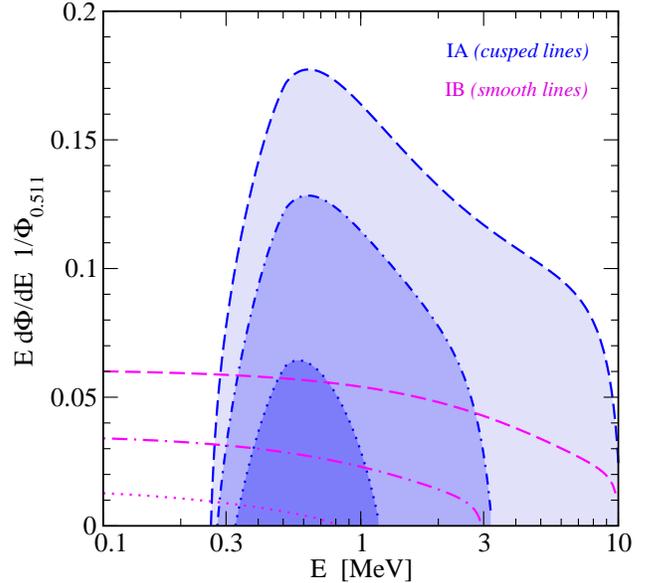}
\caption{Gamma-ray spectra (cusped lines, inflight annihilation; smooth lines, internal bremsstrahlung) from relativistic positrons, normalized to the 0.511 MeV flux; the injection energies are 1, 3, and 10~MeV (dotted, dot-dashed and dashed).
\label{fig:galpos-1}}
\end{figure}

We review the energy loss and confinement of positrons in the GC and calculate the survival probability for relativistic injected positrons to reach nonrelativistic energies.  We then calculate the IA gamma-ray spectrum; for various injection energies, Fig.~\ref{fig:galpos-1} shows the expected IA and IB spectra.  After comparing to Galactic diffuse gamma-ray data, we conclude that the IA flux is only compatible if the positrons are injected with energies $\lesssim 3$ MeV, which limits the fraction of positrons annihilating in flight to $\lesssim 5.5\%$.  The generality of our constraint on the positron injection energy allows strong discrimination between models~\cite{Models-Astro,Models-Exotic}.
 
{\bf Positron Survival Probability.---}
We assume the positrons are injected monoenergetically at a total energy $E_0$, and lose energy  while remaining confined to the GC by magnetic fields (with net displacements $\lesssim 100$ pc, much less than the $\sim 1$ kpc size of the emission region~\cite{Jean05}).  The energy loss rate for a positron of energy $E$ due to ionization in a neutral hydrogen medium of number density $n_{\rm H}$ is (see Ref.~\cite{energyloss} for the more exact form used in our calculations)
\begin{equation}
\Big|\frac{dE}{dx} \Big| \simeq \frac{7.6 \times 10^{-26}}{\beta^2}
\bigg(\frac{n_{\rm H}}{0.1 \;\rm{cm}^{-3}}\bigg) (\ln \gamma +6.6) \;
\frac{\rm MeV}{\rm cm},
\label{eq:energyloss}
\end{equation}
where $\gamma = E/m_e = 1/\sqrt{1 - \beta^2}$ is the Lorentz factor and $\beta$ is the velocity.  For $E \lesssim 100$ MeV, the in-medium bremsstrahlung and inverse Compton energy losses can be neglected~\cite{energyloss}. If the medium is fully ionized, the energy loss rate (Coulomb scattering instead of ionization) will be about 3 times larger~\cite{energyloss}.  While it has been suggested that the INTEGRAL positron data alone may favor either a single-phase weakly-ionized warm medium~\cite{Churazov04} or a multi-phase medium dominated by warm neutral and warm ionized phases~\cite{Jean05}, we assume a medium of neutral hydrogen since direct astronomical probes suggest~\cite{Moskalenko:2001ya} a weak ionized component ($\sim 10\%$) at the GC, which would only mildly affect our results.  

Dirac was the first to calculate the annihilation cross section of positrons on electrons at rest~\cite{Dirac}; it is
\begin{equation}
\sigma = \frac{\pi r_e^2}{\gamma + 1}
\left[ \frac{\gamma^2+4\gamma+1}{\gamma^2-1}
\ln{(\gamma+\sqrt{\gamma^2-1})}
- \frac{\gamma+3}{\sqrt{\gamma^2-1}} \right],
\label{eq:totcross}
\end{equation}
where $r_e$ is the classical electron radius.  (The electrons at the GC can be assumed to be at rest~\cite{Churazov04, Jean05, Guessoum05}; when the electron motion must be taken into account, see Refs.~\cite{Svensson, Moskalenko}.)  The energy loss is the same for every positron traversing the same distance (more accurately, the same column density). If we define the mean number of positrons with energy between ($E$, $E+dE$) as $N(E)$~\cite{Furlanetto}, then the fraction of positrons annihilating as they travel a distance $dx$ and lose an energy $dE$ is
\begin{equation}
\frac{dN_{}(E)}{N_{}(E)}= n_{\rm H} \, \sigma(E) \, dx = n_{\rm H} \,
\sigma(E) \, \frac{dE}{| dE/dx |},
\label{eq:difsurvival}
\end{equation}
where $n_{\rm H}$ is equal to the number density of electrons (bound plus free).  The integrated survival probability of positrons as they lose energy from $E_{\rm 0}$ to $E$ is~\cite{Murphy}
\begin{equation}
{\rm P}_{E_0\rightarrow E} = \frac{N_{}(E)}{N_{}(E_0)} =
\exp\left( - n_{\rm H}\int_{E}^{E_0} \sigma( E')\, \frac{ d E'} {| dE'/dx |}\right),
\label{eq:survival}
\end{equation}
The density dependence cancels since $| dE/dx |$ scales with $n_{\rm H}$.  When the positrons have lost most of their energy, $E \simeq m_e$; we use ${\rm P} = {\rm P}_{E_0\rightarrow m_e}$ for this terminal survival probability.  The argument of the exponential is small: for injection energies of 10 (3, 1) MeV, P differs from unity by $\simeq$ 11 (5.5, 1.4) \%.  The dominant contributions to IA are at increasingly lower energies, as the cross section is rising, until the energy loss rises rapidly when the positrons become nonrelativistic.  The rate of nonrelativistic positron annihilation in the GC, $\dot N_{}(m_e) \sim  10^{50}$ year$^{-1}$, has been measured by INTEGRAL.  Assuming equilibrium of the injection and annihilation rates, the small IA fraction increases the required positron injection rate, so that $\dot N_{}(E_0) = \dot N_{}(m_e)/ {\rm P}$. 

Nonrelativistic positrons may either directly annihilate with an electron, producing two 0.511 MeV gamma rays or, due to the low temperature of the interstellar medium, form a positronium bound state with an electron~\cite{Steigman, Stecker, Leventhal}.  Positronium annihilates to two gamma rays (each 0.511 MeV) 25\% of the time, and to three gamma rays (each less than 0.511 MeV) 75\% of the time.  In the Galaxy, the relative intensities of the three-gamma continuum and two-gamma line emission fix the  positronium fraction to be $f = 0.967 \pm 0.022$~\cite{Jean05}.  The number of gamma rays contributing to the 0.511 MeV line per annihilated nonrelativistic positron is $2 (1 - f) + 2 f 1/4 = 2 (1 - 3 f / 4)$, so that the true annihilation rate is 3.6 times {\it larger} than would be deduced from the 0.511 MeV flux alone.  The IA of energetic positrons produces two gamma rays, and so the ratio of the total flux of IA to 0.511 MeV gamma rays is
\begin{equation}
\frac{\Phi_{\rm IA}}{\Phi_{0.511}}
=\frac{2 \;(1-{\rm P})}{2 \;(1 - 3 f / 4) \; {\rm P} }
=\frac{1}{1 - 3 f / 4}\; \frac{1-{\rm P}}{{\rm P}}.
\label{eq:ratios}
\end{equation}
Since the 0.511 MeV and IA gamma rays are both emitted isotropically, we normalize our results to the observed rate and angular distribution of the 0.511 MeV data, eliminating the need for detailed modeling of the sources and the positron propagation.


{\bf Inflight Annihilation Spectra.---}
The angle-averaged differential cross section~\cite{Stecker, Svensson, Aharonian00} for IA, in terms of the scaled gamma-ray energy $k = E_\gamma / m_e$, is 
\begin{equation}
\frac{d\sigma}{dk} = \frac{\pi r_{\rm e}^2}{\gamma^2\beta^2}
\left( \frac{-(3+\gamma)/(1+\gamma)+(3+\gamma)/k-1/k^2}
{[1-k/(1+\gamma)]^2}-2 \right)
\label{eq:diffcross}
\end{equation}
where $\gamma (1 - \beta) \leq 2k - 1 \leq \gamma (1 + \beta)$.  This formula is weighted with the gamma-ray multiplicity of 2 and thus integrates to twice the total cross section given in Eq.~(\ref{eq:totcross}).  Since the electrons are at rest, the differential cross section is sharply peaked at the endpoints of the $k$ range.  As injected positrons simultaneously lose energy and are annihilated, the low-energy peak remains below 0.511 MeV, where gamma rays are accumulated, while the high-energy peak moves slowly down to 0.511 MeV, producing a long tail.  Combining Eqns.~(\ref{eq:difsurvival}--\ref{eq:diffcross}) yields the shape of the integrated gamma-ray spectrum produced by the IA of positrons as they lose energy:
\begin{equation}
\frac{d\Phi_{\rm IA}}{dk} = \frac{\Phi_{\rm 0.511}}{1 - 3 f / 4} \; \frac{n_{\rm H}}{\rm P}  
\int^{E_0}
{\rm P}_{E_0\rightarrow E} \; \frac{1}{2} \frac{d\sigma}{dk} \; \frac{dE}{| dE/dx |}.
\label{eq:IAspectrum}
\end{equation}
The lower limit of the integral is dictated by $k$, due to energy considerations~\cite{Stecker}.  In Fig.~\ref{fig:galpos-1}, we show the IA gamma ray spectra for various positron injection energies.  The choice of plotting $E \, d\Phi/dE$ means that the variations in height directly reveal the variations in the numbers of gamma rays in each logarithmic energy interval: the area under each IA curve is proportional to twice the fraction of positrons annihilated in flight, as  given by Eq.~(\ref{eq:ratios}).  The IA flux below 0.511 MeV is small compared to the three-gamma continuum.  The gamma rays above 0.511 MeV are spread over a broad energy range, but their detectability is greatly enhanced since (a) there are no competing signals from nonrelativistic positron annihilation, and (b) the Galactic diffuse gamma-ray background is steeply falling with energy.  We also show the IB gamma-ray spectrum, assuming that the positrons are co-produced with equal-energy electrons, which doubles the IB signal from positrons alone~\cite{Beacom04}.


{\bf INTEGRAL \& COMPTEL Constraints.---}
The 0.511 MeV flux from the GC was measured by INTEGRAL to be $\left(1.07 \pm 0.03 \right) \times 10^{-3}$ photons cm$^{-2}$ s$^{-1}$~\cite{Jean05}.  The angular distribution is a two-dimensional Gaussian of $\simeq 8^\circ$ FWHM (and $\sigma \simeq 3.4^\circ$); this is confirmed by the positronium emission below 0.511 MeV~\cite{Weidenspointner06}.  This gives a 0.511 MeV peak flux of $\simeq$ 0.048 photons cm$^{-2}$ s$^{-1}$ sr$^{-1}$, with $\simeq$ $24\%$ ($80\%$) of the photons coming from a circle of half-angle $2.5^\circ$ ($6^\circ$); the corresponding average fluxes are $\simeq$ $0.042$ ($0.025$) photons cm$^{-2}$ s$^{-1}$ sr$^{-1}$. 

The flux within a circle aligned with the GC will have two components.  First, the calculated gamma-ray spectrum from IA (and IB) for an assumed injection energy; we minimize the model dependence by normalizing this flux and angular distribution to the observed 0.511 MeV data.  (The IA and IB fluxes are concentrated just in this circle.)   Second, the diffuse gamma-ray background; this is slowly varying (showing no excess at the GC like that for the 0.511 MeV line), and we normalize it from the measured INTEGRAL and COMPTEL data averaged in the inner Galactic Plane.  If this combined flux at the GC is incompatible with the diffuse data alone, then the assumed positron injection energy is too large, and is disallowed.   For monoenergetic injections, there would also be a sharp step in the energy spectrum inside the GC circle.  We assume that the rates of positron production, energy loss, and annihilation are nearly in equilibrium when averaged over the propagation timescale (e.g., $\lesssim 3.5$ Myr for $\lesssim 3$ MeV).  If positrons are not fully confined to the GC by magnetic fields, then the required positron production rate and the IA and IB fluxes will be even higher.

\begin{figure}[t]
\includegraphics[width=3.25in,clip=true]{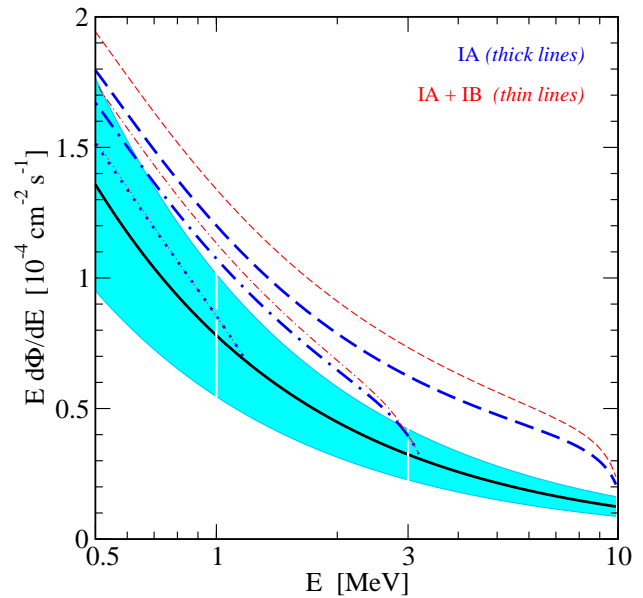}
\caption{The INTEGRAL and COMPTEL diffuse gamma ray flux measurements are shown with a black solid line, and their $\pm$30\% uncertainties by the shaded band.  For positron injection energies of 1, 3, and 10 MeV (dotted, dot-dashed and dashed lines), the thick lines show how this would be {\it increased} by the inflight annihilation gamma ray flux (thin lines also include the internal bremsstrahlung flux).
All results are for a $5^{\circ}$-diameter region at the Galactic Center.  The 0.511 MeV line flux is not shown.
\label{fig:galpos-2}}
\end{figure}

For the diffuse flux, we used the power law 
\begin{equation}
\frac{d\Phi}{dE} = 0.013 \left( \frac{E}{{\rm MeV}} \right)^{-1.8} \;
{\rm cm^{-2}\ s^{-1}\ sr^{-1}\ MeV^{-1}}.
\label{eq:powlaw}
\end{equation}
This reproduces the COMPTEL measurement of 0.0096 cm$^{-2}$ s$^{-1}$ sr$^{-1}$ between 1--3~MeV (0.0043 between 3--10 MeV) in the inner Galactic Plane ($330^\circ < l <30^\circ$, $|b| < 5^\circ$) and also agrees reasonably well with the power law determination below 1~MeV (with the positronium three-gamma continuum subtracted) from the INTEGRAL data alone in the region $350^\circ< l<10^\circ$, $|b| < 10^\circ$~\cite{Strong98,Strong05, Strong-pc}.   In Fig.~\ref{fig:galpos-2}, we show the diffuse spectrum, obtained by scaling the data averaged over this region with the solid angle of our $5^{\circ}$-diameter circle at the GC; the generous $\pm$30\% uncertainties are shown as a shaded band (primarily a systematic uncertainty on the normalization, due to subtracting detector backgrounds).

In Fig.~\ref{fig:galpos-2}, we show how our predictions for the IA flux would increase the average diffuse flux in the same GC circle.  Injection energies $\gtrsim 10$ MeV are clearly disallowed since the IA flux would more than double the diffuse flux in this circle, giving a huge spike.  Injection energies $\gtrsim 3$ MeV would cause a significant enhancement in the 1--3 MeV flux in the GC circle, relative to adjoining regions.  (Injection energies $\gtrsim 5$ MeV give a significant enhancement in both the 1--3 and 3--10 MeV fluxes.)  This argument can be strengthened by using the COMPTEL diffuse skymaps directly, instead of just the averaged data shown in Fig.~\ref{fig:galpos-2}.  We compare to the measured flux in strips of $5^\circ$ longitude and $10^\circ$ latitude~\cite{Strong98}.  The 1--3 and 3--10 MeV skymaps both show a moderate dip and peak structure (lower and higher flux) in strips straddling the GC, suggesting that its origin is in the diffuse gamma-ray background, and unrelated to positrons.  If positrons are injected at $\gtrsim 3$ MeV, then the IA flux at 1--3 MeV in the dip strip would be quite significant compared to the conventional diffuse flux there.

Thus, even with this simplified but conservative analysis, we can conclude that the positron injection energies must be $\lesssim 3$ MeV; higher energies are very strongly excluded.  Including IB would strengthen our conclusions.  Interestingly, the INTEGRAL diffuse data do show some excess in the 0.5--1.0 MeV flux in the GC relative to the adjoining regions~\cite{Strong05}.  While not strong enough to claim discovery of the IA flux, it is an intriguing hint, highlighting the importance of a more sophisticated analysis, with optimized binning of the data in energy and angle.


{\bf Conclusions.---}
The origins of the Galactic positrons remain mysterious.  We have shown that recent high-quality data from INTEGRAL on the flux and angular distribution of the 0.511~MeV gamma-ray line enable a model-independent test of the origins of the positrons.  These data allow us to accurately calculate the gamma-ray spectra from IA and IB produced by the positrons {\it while} they are relativistic, and to compare this signal to measurements of the diffuse gamma-ray flux.  Our results directly probe the positron injection energy, requiring it to be $\lesssim 3$ MeV, and apply to any positron production mechanism.  Since this is far below the energy scales suggested in some recent astrophysical~\cite{Models-Astro} and exotic~\cite{Models-Exotic} models, this is quite constraining.

Our results complement detailed studies of nonrelativistic positron annihilation, and the prospects for a solution obtained with a comprehensive approach to those, the diffuse backgrounds, and the IA and IB fluxes are excellent.  Since we can already easily probe injection energies as small as 3 MeV, very close to the energy scale of nuclear beta decays from fresh nucleosynthesis products, {\it this will very likely allow a first detection of the positron inflight annihilation signal}, and INTEGRAL data appear to show a hint already.  The angular map of positron injection positions given by IA and IB emission should more faithfully reveal the sources, in addition to the information from the injection energy scale alone.  The mismatch between this map and the 0.511 MeV map would provide new insights on Galactic magnetic fields and the conditions of the interstellar medium.


We thank Andreas Piepke, Gary Steigman, Paul Vetter, Georg Weidenspointner, and especially Andrew Strong for discussions.  This work was supported by The Ohio State University and by NSF CAREER grant No.~PHY-0547102 to J.F.B.




\end{document}